\documentclass[10pt,journal]{IEEEtran}
\usepackage{cite}
\usepackage{graphicx}
\usepackage{epsfig}
\usepackage{psfrag}

\newcommand{\cA}{{\mathcal{A}}}

\newcommand{\cE}{{\mathcal{E}}}

\newcommand{\CN}{{\mathcal{CN}}}

\newcommand{\cP}{{\mathcal{P}}}

\newcommand{\cS}{{\mathcal{S}}}

\newcommand{\cT}{{\mathcal{T}}}

\newcommand{\cY}{{\mathcal{Y}}}

\newcommand{\al}{\alpha}

\newcommand{\g}{\gamma}

\newcommand{\eps}{\varepsilon}

\usepackage[tbtags]{amsmath} 
\usepackage{verbatim} 
\newcounter{actr}
{\begin{list}{(\alph{actr})}{\usecounter{actr}}}{\end{list}}

\newcounter{ictr}
{\begin{list}{(\roman{ictr})}{\usecounter{ictr}}}{\end{list}}

\newtheorem{thm}{Theorem}
\newtheorem{lemma}{Lemma}

\newtheorem{prop}{Proposition}
\newtheorem{defn}{Definition}

\newenvironment{new-proof}[1]
{{\em Proof of #1: }}%
{ \noindent\qed }
\newcommand{\qed}{\rule[0.1ex]{1.4ex}{1.6ex}}

\newcommand{\defeq}{\stackrel{\Delta}{=}}

\newcommand{\compls}{\mathbb{C}}

\newcommand{\mrm}{\mathrm}

\DeclareMathOperator{\diag}{diag}
\usepackage{amssymb}
\newcommand{\NS}{{\mathrm{ns}}}
\usepackage[mathcal]{euscript}  

\title{Fundamental Limits and Scaling Behavior of Cooperative
Multicasting in Wireless Networks}

\author{Ashish~Khisti,~\IEEEmembership{Student~Member,~IEEE,}
        Uri~Erez,~\IEEEmembership{Member,~IEEE,}\\
        Gregory~W.~Wornell,~\IEEEmembership{Fellow,~IEEE}%
        \thanks{Manuscript received March 14, 2005; revised November
        29, 2005.  This work was supported in part by the National
        Science Foundation under Grant No.~CNS-0434974, by
        Hewlett-Packard through the MIT/HP Alliance, and by NEC
        Corporation.  This work was presented in part at the 2004
        Allerton Conference on Communication, Control and Computing.}
        \thanks{A.~Khisti and G.~W.~Wornell are with the Department
        of. Electrical Engineering and Computer Science, Massachusetts
        Institute of Technology (MIT), Cambridge, MA, 02139, USA
        (E-mail: \{khisti,gww\}@mit.edu).  U.~Erez is with the
        Department of Electrical Engineering-Systems, Tel Aviv
        University, Ramat Aviv, 69978, Israel (E-mail:
        uri@eng.tau.ac.il). }}

\begin{document}


\maketitle

\begin{abstract}

A framework is developed for analyzing capacity gains from user
cooperation in slow fading wireless networks when the number of nodes
(network size) is large.  The framework is illustrated for the case of
a simple multipath-rich Rayleigh fading channel model.  Both
unicasting (one source and one destination) and multicasting (one
source and several destinations) scenarios are considered.  We
introduce a meaningful notion of Shannon capacity for such systems,
evaluate this capacity as a function of signal-to-noise ratio (SNR),
and develop a simple two-phase cooperative network protocol that
achieves it.  We observe that the resulting capacity is the same for
both unicasting and multicasting, but show that the network size
required to achieve any target error probability is smaller for
unicasting than for multicasting.  Finally, we introduce the notion of
a network ``scaling exponent'' to quantify the rate of decay of error
probability with network size as a function of the targeted fraction
of the capacity.  This exponent provides additional insights to system
designers by enabling a finer grain comparison of candidate
cooperative transmission protocols in even moderately sized networks.

\end{abstract}


\section{Introduction}

\PARstart{C}{ooperative} diversity has been proposed as an attractive
approach to combatting slow fading in wireless networks
\cite{SendonarisErkipAazhang03, LanemanTseWornell04}.  Spatially
distributed nodes provide an opportunity to create a distributed
virtual antenna array and can provide substantial gains in slow fading
environments. There has been a significant interest in studying these
gains recently; see, e.g., \cite{LanemanWornell03, AzarianElGamal05,
NabarBolcskeiKneubuhler04, PrasadVaranasi04, BletsasKhisti04} and the
references therein.

A convenient channel model for such problems, as has been widely
adopted in the literature, is a quasistatic one in which the
parameters are known to receivers, but not to transmitters.  In such
scenarios the classical Shannon capacity is typically zero due to the
positive probability of the channel experiencing an arbitrarily deep
fade, so performance is instead quantified in terms of outage
capacity, which describes the achievable rate subject to a constraint
on the level of outage probability that can be tolerated
\cite{OzarowShamaiWyner94}.

Outage analysis applies to a host of multiterminal extensions of such
basic channel models as well \cite{Teletar99}, although the
expressions become more cumbersome.  To address this,
diversity-multiplexing tradeoff analysis provides a suitably coarser
scale characterization of such systems by focusing on the high
signal-to-noise ratio (SNR) regime and examining how outage
probability scales with SNR in this regime for different transmission
rates \cite{ZhengTse03}.

Diversity-multiplexing tradeoff analysis has also proven useful
analyzing a host of simple network problems.  For example,
\cite{TseViswanathZheng04} extends the analysis to the multiple-access
channel, while \cite{LanemanTseWornell04} extends the analysis to the
cooperative diversity channel.

While such analysis of cooperative diversity has proven popular, much
of the work has been limited to systems in which for a given message
there is effectively only a single destination node and a relatively
small number of potential relay nodes to participate in the
transmission.

In the present paper, we develop an alternative framework within which
to examine cooperative protocols.  First, our emphasis is on the
multicasting scenario in which there is one message in the network,
but generally multiple destination nodes.  We will focus on two
extreme special cases of this scenario.  One is when \emph{all} nodes
in the network are to receive the message, which for convenience we
generically refer to as multicasting.  The other is when exactly one
node in the network is to receive the message, which we refer to as
unicasting.

Second, our framework examines the scenario in which the number of
nodes in the network is large.  This will allow us to examine the
associated asymptotic scaling behavior of cooperative networks.  As a
by-product, we do not need to restrict our attention to high SNR
analysis.  Indeed, we fix the noise power, normalize the channel
statistics, and contrain the total power transmitted in the network.
This allows us to parameterize our results in terms of the associated
SNR.

Within this framework, we analyze the relationship between
transmission rate and the associated error probability.  Provided we
use codes of sufficiently long block lengths, outage probability
dominates the error probability.  Specifically, the associated outage
event is that not \emph{all} the intended recipients are able to
decode the message.  Not suprisingly, avoiding outage in multicasting
is more difficult than in unicasting.

As our main result, we show that under a multipath-rich Rayleigh
fading network model, a notion of Shannon capacity can be developed.
Specifically, there exists a nonzero capacity (dependent on SNR) such
that for all rates below capacity, the error probability can be made
arbitrarily small provided the network is sufficiently large.
Conversely, for all rates above capacity, the error probability is
bounded away from zero regardless of network size.  Our achievability
result is based on a simple two-phase cooperative network protocol we
develop.  By contrast, when one precludes the possibility of
cooperation, the associated capacity is of course zero.
Interestingly, our analysis also reveals that despite the fact that
multicasting outage behavior is dominated by the \emph{worst} node,
the multicasting and unicasting capacities thus defined are identical.
Not surprisingly, we also show that for a fixed number of nodes, the
probability of error is still much smaller in unicasting than in
multicasting.

We further show that finer scale characterizations of behavior are
possible too.  In particular, we define a notion of network scaling
exponent that characterizes the rate of decay of error probability
with network size as a function of the targeted fraction of capacity.
Within this analysis, we see, among other insights, that the exponent
of our capacity-achieving protocol is quite small for rates that
exceed half the network capacity.

While our results are specific to our multipath-rich Rayleigh fading
model, we believe that the associated framework is useful more broadly
in the analysis of user cooperation gains in large networks with more
realistic --- if more complicated --- models.  Indeed, ultimately our
results more generally suggest that just as system analysis asymptotic
in block length or SNR has proven useful, so can one that is
asymptotic in network size.

The remainder of the paper is organized as follows.
Section~\ref{sec:channel} introduces the network model of
interest. The capacity result is stated in
Section~\ref{sec:codthm}. It is established by providing a converse in
Section~\ref{sec:SimpleUpperBound} and an achievability argument in
Section~\ref{sec:ucp}.  The scaling of the outage probability with the
number of nodes is discussed in Section~\ref{sec:scaling}, and the
network scaling exponent is introduced in
Section~\ref{Network-Scaling-Exponent}.  Finally,
Section~\ref{sec:Conclusion and Future Work} contains some concluding
remarks and directions for future work.

\section{System Model}
\label{sec:channel}

We consider a system with $K$ receiving nodes and one source node.
For convenience, we label the source node as node 0, and the receiving
nodes as $\{1,2,\dots, K\}$.  Of course, in practice, different nodes
in the network can act as source nodes over orthogonal time or
frequency bands as discussed in~\cite{LanemanTseWornell04}.  However,
for analysis, it suffices to focus on a single configuration.

We assume a narrowband, slow fading channel passband channel model
corresponding to a multipath-rich propagation environment.  In
particular, the channel gains $h_{jk}$ between arbitrary distinct
pairs of nodes $(j,k)$ are independent identically distributed
(i.i.d.) random variables from a zero-mean, unit-variance circularly
symmetric complex Gaussian distribution. In turn, the signal received
at node $k$ at time $i$ is given by
\begin{equation}
\label{eq:Ch_model1}
y_k(i) = \sum_{j \in \cT(i)} h_{jk}\, x_j(i),
+ z_k(i),
\end{equation}
where $\cT(i)$ is the set of nodes transmitting at time $i$, where
$x_j(i)$ is the symbol transmitted by node $j$ at time $i$, and where
$z_k(i)$ denotes circularly symmetric complex i.i.d.\ Gaussian noise
of power $N_0$. Furthermore, the noises among the different receivers
are mutually independent.

In our model, nodes are subject to a half-duplex constraint, i.e., a
node cannot transmit and receive simultaneously.  Thus, associated
with every valid protocol is a set of binary variables of
the form $D_k(i)$ that specifies at time $i$ whether node $k$ is
transmitting ($D_k(i)=1$) or receiving ($D_k(i)=0$).

The source sends one of $M$ possible messages to the destination
node(s) over $n$ channel uses (i.e, $i=1,2,\dots,n$).  The channel
gains between all pairs of nodes remains fixed over this duration.
In our model, the channel gain $h_{jk}$ is known to the receiving
node $k$ but not to the transmitting node $j$.

We further restrict our attention to protocols in which relay nodes
cannot revert to receive mode once they begin transmitting, i.e., if
$D_k(i)=1$ for some node $k$ and $i < n$ then $D_k(j)=1$ for all
$i\leq j \leq n$.  This restriction precludes protocols in which
transmitting nodes effectively learn and exploit the network channel
gains in their encodings.

Finally, for simplicity, we adopt a long-term sum power constraint
across the nodes in our model.  In particular, with $X_j(i)$ for $j\in
\cT(i)$ denoting the (complex-valued) symbol being transmitted by node
$j$ at time $i$, we impose an expected sum power constraint of the form
\begin{equation}
\label{eq:sum_power_const}
E\left[\frac{1}{n}\sum_{i=1}^n\sum_{k\in \cT(i)}|X_k(i)|^2\right]
\leq P,
\end{equation}
where the expectation is taken over the ensemble of channel
realizations and the set $\cT(i)$, as well as any other randomized
aspects of the protocol.  Indeed, one will want to consider protocols
in which the set $\cT(i)$ depends on the realized channel gains.

The power constraint \eqref{eq:sum_power_const} is a rather natural
one for systems in which there are ergodic channel variations but a
stringent delay constraint that requires transmission of any
particular message within a single coherence interval.  Indeed,
although we do not satisfy the sum power constraint during the
transmission of an individual message, the expectation in
\eqref{eq:sum_power_const} ensures that it will be satisfied with high
probability over a sufficiently long sequence of messages.
Nevertheless, we remark in advance that the results of the paper do
not change when the we require the sum power constraint to be met with
high probability in every coherence interval, and our
capacity-achieving protocol can be readily extended to this case.
Ultimately, the expected power constraint merely simplifies the
exposition.

The preceding discussion characterizes an admissible protocol for
our analysis, which we formalize in the following definition.
\begin{defn}
\label{defn:FeasibleProtocol}
An admissible protocol $\pi_K$ consists of a set of indicator
functions $\{D_k(i)\}\in\{0,1\}$, which determines whether node $k$ is
transmitting or receiving at time $i$; a set of encoding functions
$\{\phi_k(i)\}\in\compls$, which determines the symbol produced by
node $k$ at time $i$; and a set of decoding functions
$\{\psi_k\}\in\{1,2,\dots,M\}$, which determines the message decisions
produced by node $k$ at time $n$.  These functions are further
constrained by their usage as described below.

During the initialization phase of the protocol, the source node 0
selects message $W\in\{1,2,\dots,M\}$ for transmission.  The protocol
chooses \emph{a priori} the sequence $D_0(i)$ for $i=1,2,\dots, n$.
Without loss of generality, $D_0(1)=1$ and $D_k(1)=0$ for
$k=1,2,\dots,K$.   Moreover, the collections of observations $\cY_k$
at each node $k$ are initialized:  $\cY_k=\emptyset$.

At time $i$, for $1\le i < n$, if $D_0(i)=1$, then source node 0 uses
encoding function $\phi_0(i)$ to map $W$ into a transmitted
complex-valued symbol $x_0(i)$.

If node $k \in \{1,2,\dots,K\}$ is in transmit mode at that time
(i.e., $D_k(i)=1$), the encoding function $\phi_k(i)$ at node $k$ maps
$\cY_k$ and the complex-valued channel gains $\{h_{jk},
j=0,1,\dots,K\}$ into a transmitted complex-valued symbol $x_k(i)$,
which is transmitted over the channel.

If, instead, the node is in receive mode (i.e., $D_k(i)=0$), then it
collects the complex-valued measurement $y_k(i)$ and updates its set
of received symbols via $\cY_k := \cY_k \cup \{ y_k(i) \}$.  If $i=n$,
the decoding function $\psi_k$ at node $k$ maps $\cY_k$ and the
complex-valued channel gains $\{h_{jk}, j=0,1,\dots,K\}$ into a
decision $\hat{W}_k$. Note that without loss of generality, $D_k(n)=0$
for at least one value of $k$.  If $i<n$, the node makes
a decision whether to switch to transmit mode for the remaining
duration.  If it decides to switch, it sets $D_k(j)=1$ for $i+1\le
j\le n$; otherwise, it sets $D_k(i+1)=0$.
\end{defn}

A cooperation-free protocol is a special case of the above definition:
\begin{defn}
\label{def:cf} A \emph{cooperation-free} admissible protocol is
one for which only the source node transmits, i.e., $D_k(i)=0$ for
$1\le k\le K$ and $1\le i \le n$.
\end{defn}

\section{Coding Theorems}
\label{sec:codthm}

We now develop the relationship between transmission rate and error
probability for such protocols, in the limit of large network sizes.

We begin with a meaningful definition of capacity.
\begin{defn}
\label{defn:CommonRate-Unicasting} A rate $R$ is \emph{achievable} for
the unicasting (respectively, multicasting) system if for every
network size $K$, there exists an admissible protocol $\pi_K$ with
$n_K$ channel uses and $M_K = 2^{n_KR}$ messages such that the
probability that the destination node (respectively, any node) fails
to decode the message approaches zero as $K\rightarrow\infty$. The
(unicasting or multicasting) \emph{capacity} $C$ is the supremum of
all achievable rates.
\end{defn}

With this definition, we have the following coding theorem, which is
our main result.
\begin{thm}
The unicasting and multicasting capacities are identical and given
by\footnote{Unless otherwise indicated, all logarithms are base 2.}
\begin{equation}
C= \log\left(1+\frac{P}{N_0}\right),
\label{eq:Capacity}
\end{equation}
where $P$ is as defined in \eqref{eq:sum_power_const} and $N_0$ is the
noise power as defined via \eqref{eq:Ch_model1}.
\label{MainThm}
\end{thm}

Before presenting our proof of this result, we note that to achieve
capacity --- indeed any nonzero rate --- requires the use of
cooperation.   Formally, we have the following result.
\begin{thm}
The capacity of cooperation-free admissible protocols for both
unicasting and multicasting is
\begin{equation}
C_\mathrm{nc}= 0
\label{eq:Capacity-nc}
\end{equation}
whenever $P$ in \eqref{eq:sum_power_const} is finite and $N_0$ in
\eqref{eq:Ch_model1} is nonzero.
\label{preMainThm}
\end{thm}

\begin{proof}
From Definition~\ref{def:cf}, a destination node must decode directly
from the source transmission.  Since the channel is a Rayleigh fading
channel with $P/N_0<\infty$, there exists a strictly positive
probability of outage --- and hence probability of error --- for every
positive transmission rate $R$.   Since these probabilities are
independent of $K$, \eqref{eq:Capacity-nc} follows.
\end{proof}

We now proceed to the proof of Theorem~\ref{MainThm}.  Since we
consider large block lengths, error probability is dominated by outage
probability.  Specifically, we consider blocks long enough that the
probability of error when there is no outage is negligible compared to
the outage probability.  Thus, in the error analysis in our proof, we
restrict our attention to outage probability.

\section{Proof of Converse Part}
\label{sec:SimpleUpperBound}

We develop a converse via a simple upper bound on the achievable
rate $R$ of Definition~\ref{defn:CommonRate-Unicasting}.  In
particular, suppose a genie conveys the message $W$ to nodes
$1,2,\dots, K-1$ and only destination node $K$ remains to be
served.  Thus, nodes $0,1,\dots,K-1$ can coordinate to send the
message to destination node $K$.  This is clearly a multiple-input
single-output (MISO) antenna system with $K$ antenna elements and
channel knowledge only at the receiver.  Thus, for a given rate
$R$, a lower bound on the outage probability for the MISO channel
is a lower bound on that for both unicasting and multicasting
systems. We let $\cE_K^\mathrm{MISO}$ denote the MISO channel
outage event.

To develop such a bound, we first note that it suffices to
restrict the input distribution to  Gaussian.

\begin{lemma}[Teletar \cite{Teletar99}]
\label{Lem:Teletar}
The outage capacity of the slow-fading MISO channel with $K$ transmit antennas in i.i.d.\ Rayleigh
fading with total power constraint  $P$ is achieved by  an input distribution with a covariance matrix
$\diag(P_1,P_2,\dots,P_K)$, where $\sum_{j=1}^K P_j = P$.
\end{lemma}

We now establish the following lemma:

\begin{lemma}
Let $\epsilon>0$ be arbitrary, and let
\begin{equation*}
R=\log\left(1+\frac{P}{N_0}\right) + \epsilon
\end{equation*}
for a $K$-antenna MISO channel.  Then the outage probability
$\Pr\{\cE_K^\mathrm{MISO}\}$ is bounded away from zero, i.e.,
\begin{equation}
\inf_K \Pr\{\cE_K^\mathrm{MISO}\} > 0.
\label{eq:ibz}
\end{equation}
\end{lemma}

\begin{proof}

Let $P^*_1,P^*_2,\ldots P^*_K$ be the power allocations that
minimize the outage event for the selected rate. The corresponding
mutual information is given by

\begin{equation}
I = \log\left(1 + \frac{1}{N_0} \sum_{i=1}^K
P^*_i|h_{iK}|^2\right) \label{I_eq}
\end{equation}

The outage event $\cE_K^\mathrm{MISO} = \left\{ R > I\right\}$ is
bounded as follows:

\begin{align}
\Pr\{\cE_K^\mrm{MISO}\} &= 1 - \Pr(I \ge R)\notag\\
&\ge 1 - \frac{E\left\{\log\left(1 + \frac{1}{N_0} \sum_{i=1}^K P^*_i|h_{iK}|^2\right) \right\}}{R}\label{tel:Markov}\\
&\ge 1 - \frac{\log\left(1 + \frac{1}{N_0} \sum_{i=1}^K E[|h_{iK}|^2]P^*_i\right)}{R}\label{tel:Jensen}\\
&=   1- \frac{\log\left(1 + \frac{1}{N_0} \sum_{i=1}^K P^*_i\right)}{R}\label{tel:Expectation}\\
&=   1 - \frac{\log\left(1 + \frac{P}{N_0}\right)}{R}  = \frac{\eps}{R}> 0 \notag.
\end{align}

In the above derivation~(\ref{tel:Markov}) follows from the Markov inequality,~(\ref{tel:Jensen})
is a consequence
of Jensen's inequality and~(\ref{tel:Expectation}) follows from the Rayleigh model $E[|h_{iK}|^2]=1, \forall i$.
Since the above result holds for all $K$,  (\ref{eq:ibz}) follows.

\end{proof}

\section{Proof of Forward Part}
\label{sec:ucp}

A simple two-phase cooperative protocol can achieve any rate below the
capacity \eqref{eq:Capacity}, as we now develop.

\subsection{A Two-Phase Cooperative Protocol}
\label{sec:protocol}

\begin{figure}[tbp]
\begin{center}
\includegraphics[width = 3.3 in]{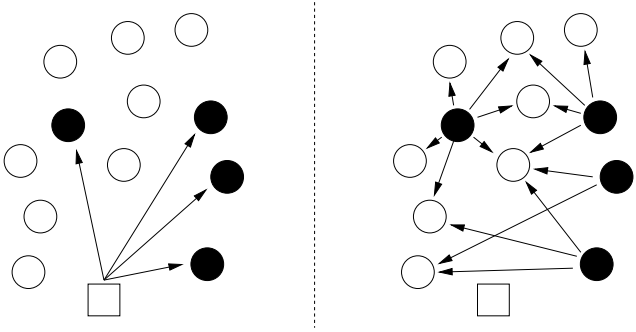}
\caption{The two-phase cooperative multicasting protocol.  In phase 1
(left), the source node (square) broadcasts at a high rate and only a
small fraction (solid) of the many destination nodes (discs) are able
to decode.  In phase 2 (right), these nodes cooperatively broadcast
the message to the remaining nodes using a suitable space-time
code.\label{fig:InnerBound}}
\end{center}
\end{figure}

The protocol of interest is depicted in Fig.~\ref{fig:InnerBound}.
Specifically, in phase 1, the source node broadcasts the message over
$n_1$ channel uses at a rate $R_1$ and all nodes attempt to decode the
message.  Then, in phase 2, the nodes that are successful in decoding
the message act as relays and form a virtual antenna array,
transmitting over the remaining $n_2=n-n_1$ channel uses at a rate
$R_2$. At this point the intended destination(s) attempt(s) to decode
the message and an outage is declared if any of the intended
destinations fail.

\paragraph*{Codebook Generation}
Suppose that the source generates $M$ codewords i.i.d.\ $\CN(0,P_1)$
for some $P_1>0$, each of length $n_1$, and all other nodes each
generate $M$ codewords i.i.d.\ $\CN(0,P_2)$ for some $P_2>0$, each of
length $n_2$.  We describe the main steps of the protocol for the case
of multicasting, but indicate the straightforward modifications for
the case of unicasting.

\paragraph*{Phase 1}
The source transmits the codeword corresponding to the intended
message from its codebook over $n_1$ channel uses.  We choose the rate
in this phase to be (strictly less than, but arbitrarily close
to\footnote{This technicality ensures that the probability of error
when not in outage will approach zero uniformly over all channel
realizations.  A similar technicality applies to the rate in phase
2.})
\begin{equation}
R_1(\alpha) = \log_2\left(1+ G(\alpha) \frac{P}{N_0}\right),
\label{eq:Ral}
\end{equation}
where
\begin{equation}
G(\alpha) = F^{-1}(1-\alpha),
\end{equation}
and $F(\cdot)$ denotes the cumulative distribution function of an
arbitrary channel gain $|h_{ij}|^2$, and where we have made the
dependency of $R_1$ on $\alpha$ explicit.

All nodes attempt to decode the transmission. A node is
successful in decoding the message if it finds a codeword in the
source codebook that is jointly typical with the received
sequence. Let $K_1$ be the number of nodes that are successful in
decoding the message from the source. We label these nodes as $1,2,
\dots, K_1$. These nodes participate in phase 2.

\paragraph*{Phase 2}
Each of the $K_1$ nodes successful in decoding the phase 1
transmission next transmits the corresponding codeword from its
codebook over $n_2$ channel uses. The rate in this phase is set to
(strictly less than, but arbitrarily close to)
\begin{equation}
\label{eq:del_defn}
R_2(\beta) = \log\left(1+\frac{P}{N_0} (1-\beta)\right),
\end{equation}
where $0<\beta<1$ is design parameter, the dependence of $R_2$ on
which we have made explicit.

Each of the remaining $K_2 = K-K_1$ nodes attempts to decode the
message at the end of the second phase.  Node $k$, upon receiving is
observations\footnote{We use the superscript $\ ^{n_2}$ to denote the
vector formed from the $n_2$ variables corresponding to time instants
$n_1+1, n_1+2, \dots, n$, i.e., phase 2 of the protocol.}  $y_k^{n_2}$
finds a message $\hat{w}_k$ and a subset of nodes $\cS_k =
\{k_1,k_2,\dots, k_{|\cS_k|}\}\subseteq \{1,2,\dots,K\}\setminus
\{k\}$ such that the corresponding set of codewords
$\{x_{k_1}^{n_2}(\hat{w}_k),x_{k_2}^{n_2}(\hat{w}_k),\dots,
x_{k_{|\cS_k|}}^{n_2}(\hat{w}_k)\}$ is jointly typical with
$y_k^{n_2}$. It declares the message $\hat{w}_k$ to be the transmitted
message if a unique pair $(\hat{w}_k,\cS_k)$ exists and declares a
failure otherwise.

In the case of unicasting, if the destination node is successful in
decoding the message in phase 1 then it does not participate in phase
2.  Otherwise it continues to listen to the transmissions and attempts
to decode the phase 2 transmission.  An error occurs if the
destination fails to decode the phase 2 transmission.

\subsection{Protocol Analysis}

First, we analyze code rate.  To begin, it is straightforward to
verify that $n_1$ and $n_2$ are completely determined by the choice of
rates.  In particular, let the overall rate of our protocol be $R$, so
that there are $M=2^{nR}$ possible messages to send over $n$ channel
uses in the system.  Then it follows that
\begin{equation}
n_1 R_1 = n_2 R_2 = \log M.
\label{eq:n1n2}
\end{equation}

From \eqref{eq:n1n2} it is straightforward to calculate the overall
effective rate $R$ of the system.  In particular, since $\log M = nR$
and since $n_1+n_2=n$, \eqref{eq:n1n2} implies that $R$ satisfies
\begin{equation}
\label{eq:Reff}
\frac{1}{R} = \frac{1}{R_1(\alpha)} + \frac{1}{R_2(\beta)}.
\end{equation}

Second, we analyze the power constraint \eqref{eq:sum_power_const}.
In phase 1, the transmitted power is $P_1$, so provided $P_1\le P$,
our power constraint is met in this phase.  To analyze the power used
in phase 2, we begin by noting that on average a fraction $\alpha$ of
the nodes are able to decode the message after phase 1.  Specifically,
the number of nodes $K_1$ successful in phase 1 has mean
\begin{equation}
E[K_1] = \alpha K
\end{equation}
since $K_1$ is a binomial random variable, viz., (cf.\ \eqref{eq:Ral})
\begin{equation}
K_1 = \sum_{i=1}^K 1_{\{|h_{0i}|^2>G(\alpha)\}},
\label{eq:K_1_defn}
\end{equation}
where $1_{\{\cdot\}}$ is an indicator function, which equals 1 if its
subscript is true and 0 otherwise, and where we have set $P_1=P$.
Hence,
\begin{equation*}
E\left[\sum_{j=1}^{K_1}|X_j|^2\right] = P_2 E[K_1] = P_2 \alpha K,
\end{equation*}
from which we see that the power constraint is satisfied in phase 2
provided $P_2 \le P/\alpha K$.

Finally, we analyze the outage probility, i.e., the probability of
outage of a node that is unable to decode at the end of the
protocol. For convenience, let us exploit symmetry and label this node
$K$, while the nodes that are successful in phase 1 we label
$1,2,\dots K_1$.  From straightforward MISO system analysis, node $K$
will fail to decode the message whenever $K_1<K$ and
\begin{equation}
R_2(\beta) \ge \log\left(1+G_K(K_1)\frac{P}{N_0}\right),
\label{eq:outage-condn}
\end{equation}
where
\begin{equation}
G_k(k_1) \defeq \frac{1}{\alpha K}\sum_{j=1}^{k_1}|h_{jk}|^2
\label{eq:G_K-defn}
\end{equation}
is the effective MISO channel gain of node $k$, and where we have
set $P_2=P/(\al K)$.  But since $R_2$ was chosen according to
\eqref{eq:del_defn} in phase 2, \eqref{eq:outage-condn} implies
that outage will occur when $G_K(K_1) \le 1-\beta$.  Accordingly,
the outage events $\cE^\mathrm{uc}_{\al,\beta,K}$ and
$\cE^\mathrm{mc}_{\al,\beta,K}$ for unicasting and multicasting,
respectively, take the form
\begin{align*}
\cE^\mathrm{uc}_{\al,\beta,K} &= \bigcup_{k_1=0}^{K-1}
\cA_{k_1}^\mathrm{uc} \\
\cE^\mathrm{mc}_{\al,\beta,K} &= \bigcup_{k_1=0}^{K-1} \cA_{k_1}^\mathrm{mc}
\end{align*}
where, for $k_1\in\{1,2,\dots,K\}$,
\begin{align*}
\cA_{k_1}^\mathrm{uc} &= \left\{ K_1=k_1,\quad G_K(k_1) \le 1-\beta
\right\} \\
\cA_{k_1}^\mathrm{mc} &= \left\{ K_1=k_1,\quad \min_{k:k_1\le k\le K}
G_k(k_1) \le 1-\beta \right\}
\end{align*}

The following lemma provides bound on the conditional outage
probability that will be useful in the sequel.
\begin{lemma}
\label{lem:cond-out}
Suppose $k_1\ge \al K(1-\beta)$ nodes are successful at the end of
phase 1. Then the conditional probability of outage is given by
\begin{equation}
\label{eq:cond_Pr_single}
\Pr\{\cE^\mathrm{uc}_{\al,\beta,K} \mid K_1=k_1\} \leq \left(\frac{\alpha
K(1-\beta)e}{k_1}\right)^{k_1} e^{-\al K(1-\beta)}.
\end{equation}
\end{lemma}

\begin{proof}
To obtain \eqref{eq:cond_Pr_single}, it suffices to bound
$\Pr\{G_K(K_1) \leq 1-\beta \mid K_1=k_1\}$ since
$\Pr\{\cE^\mathrm{uc}_{\al,\beta,K} \mid K_1=k_1\} = \Pr\{G_K(K_1)
\leq 1-\beta \mid K_1=k_1\}$.  This can be accomplished by the
Chernoff bound, since $G_K(k_1)$ in \eqref{eq:G_K-defn} can be
written as
\begin{equation}
G_K(k_1) = \sum_{j=1}^{k_1} g_j \label{eq:GK-exp}
\end{equation}
where the
\begin{equation}
g_j = \frac{1}{\alpha K}|h_{jK}|^2.
\label{eq:gj-def}
\end{equation}
are i.i.d.\ random variables.  Specifically, we obtain, for any $s>0$,
\begin{align}
&\Pr\{G_K(K_1) \leq 1-\beta \mid K_1=k_1 \} \\
&= \Pr\left\{e^{-sG_K(K_1)}\geq
e^{-s(1-\beta)} \biggm| K_1=k_1\right\} \label{eq:GkBound0}\\
&\leq e^{s(1-\beta)}E\left[e^{-sG_K(K_1)} \mid K_1=k_1\right] \label{eq:GkBound1}\\
&= e^{s(1-\beta)}\left(E[e^{-sg_i}]\right)^{k_1} \label{eq:GkBound2}\\
&=  \frac{e^{s(1-\beta)}}{\left(1+s/(\alpha K)\right)^{k_1}},
\label{eq:GkBound3}
\end{align}
where \eqref{eq:GkBound1} follows from the Markov inequality,
\eqref{eq:GkBound2} follows by from \eqref{eq:GK-exp}, and
\eqref{eq:GkBound3} follows from evaluating the characteristic
function of the exponential random variables \eqref{eq:gj-def}.

In turn, since \eqref{eq:GkBound3} holds for all $s>0$, we can choose
the particular value
\begin{equation}
s = \frac{k_1}{1-\beta}-\alpha K.
\label{eq:s-val}
\end{equation}
Substituting \eqref{eq:s-val} into \eqref{eq:GkBound3} yields
\eqref{eq:cond_Pr_single} as desired.
\end{proof}

We now show that the probability of outage can be made arbitrarily small by
selecting $\alpha K$ appropriately.
\begin{prop}
\label{prop:uc}
The probability of outage in unicasting decreases exponentially with
$\alpha K$ for every $\beta > 0$.  Specifically, for every
$0<\epsilon< \beta$,
\begin{equation}
\begin{aligned}
\Pr\{\cE^\mathrm{uc}_{\al,\beta,K}\} &=
E\left[\Pr\{\cE^\mathrm{uc}_{\al,\beta,K} | K_1\}\right] \\
&\leq \exp(-\alpha K \epsilon^2/4) + \exp\left\{\alpha K
\g(\beta,\epsilon)\right\},
\end{aligned}
\label{eq:unicast-Prob}
\end{equation}
where
\begin{equation}
\g(\beta,\epsilon) \defeq \beta-\epsilon
+(1-\epsilon)\ln\left(\frac{1-\beta}{1-\epsilon}\right)<0.
\end{equation}
\end{prop}

\begin{proof}
To obtain \eqref{eq:unicast-Prob} we observe that, for some $\epsilon
\in (0,\beta)$,
\begin{align}
&\Pr\{\cE^\mathrm{uc}_{\al,\beta,K}\} \notag\\
&= \sum_{k_1:k_1< \al K(1-\epsilon)}
\Pr\{\cE^\mathrm{uc}_{\al,\beta,K}\mid K_1 = k_1\} \Pr\{K_1 = k_1\} \notag\\
&\quad + \quad \sum_{k_1:k_1\ge \al K(1-\epsilon)}
\Pr\{\cE^\mathrm{uc}_{\al,\beta,K}\mid K_1 =k_1\} \Pr\{K_1 =k_1\} \\
&\leq
\Pr\{K_1 < \alpha K (1-\epsilon)\} +
\max_{\stackrel{k_1:}{k_1\geq \alpha K (1-\epsilon)}}
\Pr\{\cE^\mathrm{uc}_{\al,\beta,K}\mid K_1=k_1\} \notag\\
&= \Pr\{ K_1 < \alpha K (1-\epsilon)\} +
\Pr\{\cE^\mathrm{uc}_{\al,\beta,K} \mid K_1 =\alpha K(1-\epsilon)\},
\label{eq:App1-Unic-Bound-1}
\end{align}
where \eqref{eq:App1-Unic-Bound-1} exploits that outage probability is
a decreasing function of $k_1$.  Finally, using the binomial Chernoff
bound (see, e.g., \cite{Canny99})
\begin{equation*}
\Pr\{K_1 <\alpha K(1-\epsilon)\}  \leq e^{-\alpha K \epsilon^2/4},
\end{equation*}
for the first term in \eqref{eq:App1-Unic-Bound-1}, and applying
Lemma~\ref{lem:cond-out} to the second term, yields
\eqref{eq:unicast-Prob} as desired.
\end{proof}

In turn, Proposition~\ref{prop:uc} can be used to bound the
corresponding probability of outage in multicasting.
\begin{prop}
\label{prop:mc}
The probability of outage in multicasting decays asymptotically
with $\alpha K$. Specifically,
\begin{equation}
 \Pr\{\cE^\mathrm{mc}_{\al,\beta,K}\} =
E\left[\Pr\{\cE^\mathrm{mc}_{\al,\beta,K}|K_1\} \right] \leq K
\Pr\{\cE^\mathrm{uc}_{\al,\beta,K}\}. \label{eq:Multicast-Prob}
\end{equation}
\end{prop}

\begin{proof}
First, we bound the conditional outage probability according to
\begin{align}
&\Pr\{\cE^\mathrm{mc}_{\al,\beta,K}|K_1=k_1\} \\
&=
\Pr\{\min_{k:k_1\le k\le K}G_k(K_1)< 1-\beta | K_1=k_1\} \notag \\
&= \Pr\left\{\bigcup_{i=K_1+1}^K\left\{ G_{i}(K_1)
\leq 1-\beta\right\} \Biggm| K_1=k_1\right\} \notag\\
&\leq (K-k_1) \Pr\{\cE^\mathrm{uc}_{\al,\beta,K} | K_1=k_1\} \label{eq:union}\\
&\leq  K \Pr\{\cE^\mathrm{uc}_{\al,\beta,K} | K_1=k_1\}.
\label{eq:last}
\end{align}
where \eqref{eq:union} is a simple application of the union bound.
Taking the expectation of both sides of \eqref{eq:last} with respect
to $K_1$, we obtain \eqref{eq:Multicast-Prob}.   Finally, since
Proposition~\ref{prop:uc} establishes that the unicasting outage
probability decays exponentially, \eqref{eq:last} implies that the
multicasting outage probability does as well.
\end{proof}

Propositions~\ref{prop:uc} and \ref{prop:mc} can be used to
establish the forward part of the coding theorem for both
unicasting and multicasting.

\begin{new-proof}{Theorem~\ref{MainThm}}
To show that our two-phase protocol can approach the capacity
\eqref{eq:Capacity} we show that the outage probability can be made
arbitrarily small while operating arbitrarily close to the
capacity. Suppose that $\alpha>0$ and $\beta>0$ are arbitrary.  Since
the outage probability decreases exponentially in $K$, we can choose a
$K$ large enough to make the outage probability sufficiently
small. Next, note that by choosing $\alpha$ and $\beta$ sufficiently
small, we can make $R_1(\al)$ sufficiently large and $R_2(\beta)$
sufficiently close to $C$. As particular examples, it suffices to take
$\alpha \sim 1/\log K$ and $\beta\sim 1/K$ so that
$\al,\beta\rightarrow0$ but $\al K\rightarrow\infty$.  Thus, we can
have the effective rate \eqref{eq:Reff} be arbitrarily close to $C$,
while keeping the outage probability sufficiently small.

\end{new-proof}

An intuition behind the achievability result is that in the limit
of a large number of nodes, we can find sufficiently many nodes
(albeit a small fraction of the population) with very large
channel gains and they can be served over a small number of
channel uses in the first phase (i.e., $n_1$ is a negligible
fraction of $n_2$). These nodes then simultaneously cooperate to
serve the remaining nodes. Since sufficiently many nodes are
transmitting in the second phase, we have enough diversity in the
system to drive the outage probability to zero.

\subsection{Multiple Antenna Generalization}

It is possible to generalize our results to the case where the at
least some of the nodes in the network have multiple antennas. In
particular suppose that the node $i$ has $T_i$ antennas. In the
case of unicasting, our two-phase protocol can be
straightforwardly extended to obtain the following:
\begin{equation}
C = T_K\log\left(1+\frac{P}{N_0}\right). \label{eq:ma-uc}
\end{equation}
In the bound \eqref{eq:ma-uc} the key quantity of interest is $T_K$,
the number of antennas at the destination node; the number of antennas
at the source and relay nodes do not impact capacity.  Note that, in
the first phase of the protocol, we can still communicate to a large
number of relay nodes, regardless of the number of antennas at each
relay. These nodes then form a virtual antenna to communicate to the
destination in phase 2. This reduces to the case of a multiple-input
multiple-output (MIMO) system when the number of transmit antennas is
much larger than the receive antennas.  Using the channel hardening
result for such systems --- see, e.g., \cite{TarokhHochwaldMarzetta04}
--- one can establish that rate $R_\mathrm{ma}^\mathrm{uc}$ is
achievable. The converse is analogous to the single antenna case
in Section~\ref{sec:SimpleUpperBound}.

An analogous argument for multicasting can also be developed, from
which we have the following:
\begin{equation}
C = \min\{T_1,T_2,\ldots,T_K\} \cdot
\log\left(1+\frac{P}{N_0}\right). \label{eq:ma-mc}
\end{equation}
Evidently, \eqref{eq:ma-mc} can be much smaller than \eqref{eq:ma-uc}
--- the lower bound system rate is limited governed by the node with
the fewest antennas in multicasting rather than the destination node.

\section{Outage Scaling Behavior}
\label{sec:scaling}

Our capacity result determines the rates for which outage probability
goes to zero with increasing network size for multicasting and
unicasting.  Often, a finer grain analysis is required by system
designers.  In this section, we develop the manner in which outage
probability goes to zero with increasing network size for the
two-phase protocol of Section~\ref{sec:ucp}, which provides several
additional insights.  For example, while we have shown that
multicasting and unicasting share the same capacity, here we show how
their respective outage probability curves differ.

\subsection{Outage Probability Approximations}

While \eqref{eq:unicast-Prob} and \eqref{eq:Multicast-Prob} bound the
outage probabilities of interest, these bounds are not tight.
Nevertheless, good approximations to the actual outage are readily
obtained, as we now develop.

The outage probability of a unicasting system under the two-phase
protocol can be approximated by
\begin{equation}
\begin{aligned}
\Pr\{\cE^\mathrm{uc}_{\al,\beta,K}\} &\approx
\frac{1}{\sqrt{K}}{\exp\left\{-\al K(1-\beta)\right\}} \quad
\times  \\
&\exp\left\{-KD(\gamma\| \al)+
\gamma K\ln\left(\frac{\al(1-\beta)e}{\gamma}\right)\right\},
\label{eq:unicast-Prob-approx}
\end{aligned}
\end{equation}
where
\begin{equation}
\gamma = \frac{\sqrt{1+4\mu}-1}{2\mu}
\label{eq:ga}
\end{equation}
with
\begin{equation}
\mu = \frac{\alpha^2(1-\beta)}{1-\alpha}.
\label{eq:mu}
\end{equation}

In turn, the outage probability of the multicasting system can be
approximated in terms of this unicasting approximation according to
\begin{equation}
\Pr\{\cE^\mathrm{mc}_{\al,\beta,K}\} =
1-(1-\Pr\{\cE^\mathrm{uc}_{\al,\beta,K}\})^K \approx K
\Pr\{\cE^\mathrm{uc}_{\al,\beta,K}\}.
\label{eq:multicast-Prob-approx}
\end{equation}

A derivation of the approximation \eqref{eq:unicast-Prob-approx} is
provided in the Appendix.

\subsection{Accuracy of Outage Probability Approximations}
\label{sec:simulation}

In this section, we compare our outage probability bounds
\eqref{eq:unicast-Prob} and \eqref{eq:Multicast-Prob}; and our
approximations \eqref{eq:unicast-Prob-approx} and
\eqref{eq:multicast-Prob-approx}, to the actual probabilities via Monte
Carlo simulations.  In particular, we choose a target rate below
capacity and evaluate the outage probability as a function of the
network size $K$.  We evaluate the expectations over $K_1$ in the
bounds \eqref{eq:unicast-Prob} and \eqref{eq:Multicast-Prob}
by numerical integration.

For our comparison, we set a rate of
\begin{equation*}
R = \frac{1}{2}\log\left(1+\frac{P}{2N_0}\right),
\end{equation*}
which is $1/2$ of capacity in the high SNR regime and $1/4$ of
capacity in the low SNR regime.  This rate point is realized by the
parameter settings $G(\alpha)=1/2$ and $\beta=1/2$ in our two-phase
protocol, so $R_1(\alpha) = R_2(\beta)$) in \eqref{eq:Ral} and
\eqref{eq:del_defn}, respectively.

\begin{figure}[tbp]
\includegraphics[width = 3.25in]{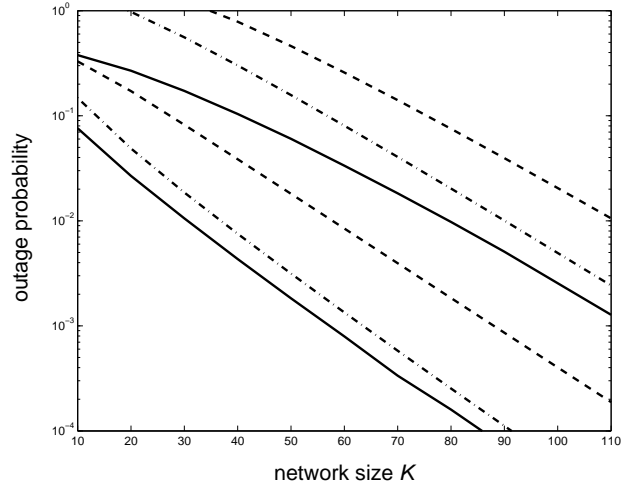}
\caption{Outage probability for unicasting and multicasting via our
  two-phase protocol as a function of network size. The solid curves
  correspond to our Monte Carlo simulations, the dashed-dotted curves
  to our analytical approximations, and the dashed curves to our
  bounds.  The top set of curves is for multicasting; the bottom set for
  unicasting.  In the protocol we set $\beta = 0.5$ and
  $R_1(\alpha)=R$, so that the rate is $R = (1/2)\log(1+P/(2N_0)) <
  C$.
\label{fig:Comparison}}
\end{figure}

Fig.~\ref{fig:Comparison} depicts the results. Several observations
are worth emphasizing.

\subsubsection*{Remarks}

\begin{enumerate}

\item First, the outage curves for both unicasting and multicasting
approach zero with our cooperative protocol, which is a consequence
of the transmission rate being below capacity.  Note that, by
contrast, for cooperation-free admissible protocols, the outage curves
will not decay with network size.

\item Multicasting incurs significant penalty over unicasting in terms
of outage probability for a fixed network size $K$.  In particular,
Fig.~\ref{fig:Comparison} confirms that the multicasting outage
probability is indeed roughly a factor $K$ larger than the unicasting
outage probability.

\item The slopes of the outage log-probability curves are
asymptotically constant, and the bounds are good predictors of the
asymptotic slopes.  This is perhaps not surprising since we used
Chernoff techniques to derive the bounds.  Indeed, in many
communication problems the Chernoff exponent is close to the correct
exponent.  However, the bounds are not particularly close to the the
correct outage curves.

\item The analytical outage probability approximations are
asymptotically quite close to the true curves, converging to within a
factor of roughly 3 in probability for large network sizes.  In
addition, these approximations appear to be actual upper bounds at
least in case study depicted, though this is conjecture.

\item The asymptotic slopes of the outage log-probability curves for
both unicasting and multicasting are identical.  In the next section,
we will develop this slope as the network scaling exponent of the
protocol, which we denote using $E_\mathrm{ns}^-$.  For a target
outage level, this slope can be used to quantify the asymptotic
network size gap between unicasting and multicasting.  In particular,
suppose that for a fixed choice of $\al$ and $\beta$ in the protocol,
$K^\mathrm{uc}(\epsilon)$ nodes are required to achieve some target
outage probability $\epsilon$ in unicasting. Then the number of
nodes required to achieve the same outage probability in multicasting
is, asymptotically,
\begin{equation}
K^\mathrm{mc}(\epsilon) = K^\mathrm{uc}(\epsilon) +
\frac{1}{E_{\NS}^-}\log K^\mathrm{uc}(\epsilon).
\label{eq:KMKU}
\end{equation}
To verify \eqref{eq:KMKU}, it suffices to recognize that the vertical
distance between the unicasting and multicasting outage probabilities
is, in accordance with \eqref{eq:multicast-Prob-approx},
asymptotically, $\log K^\mathrm{uc}(\epsilon)$.

\end{enumerate}

\section{Network Scaling Exponent}
\label{Network-Scaling-Exponent}

In this section, we explore, in more detail, the asymptotic rate of
decay of the outage probability with network size, which we have
termed the network scaling exponent.  This exponent captures
meaningful information for system designers.  For example, at the
transmission rates to which Fig.~\ref{fig:Comparison} corresponds,
outage probabilities for our two-phase protocol decay reasonably
quickly in a practical sense --- i.e., the network scaling exponent is
reasonably large.  However, as we will see, at rates close to
capacity, it turns out that outage probabilities decay very slowly as
a function of network size, corresponding to a small network scaling
exponent.  This implies that very large network sizes may be needed to
achieve practical target error rates.

Before beginning our development, note that the network scaling
exponent is the natural counterpart to the classical error exponent
for traditional channel codes.  In particular, the classical error
exponent captures the exponential rate of decay of error probability
with block length as a function of the the targeted fraction of
capacity; see, e.g., \cite{Gallager68}.  Analogously, the network
error exponent captures the exponential rate of decay of error
probability in unicasting and multicasting with network size as a
function of the targeted fraction of capacity.

Formal definitions follow.
\begin{defn}
The \emph{network reliability function} with respect to a sequence
of admissible protocols $\pi_K$ in
Definition~\ref{defn:FeasibleProtocol} is given by
\begin{equation}
E_{\NS}^-(\{\pi_K\}) = -\lim_{K\rightarrow \infty} \frac{\ln
\Pr\{\cE_{\pi_K}\}}{K} \label{eq:Reliab-exp},
\end{equation}
where $\cE_{\pi_K}$ denotes the outage event for a system with $K$
nodes under the protocol $\pi_K$.
\end{defn}

\begin{defn}
The \emph{network scaling exponent} is the supremum of the network
reliability functions of all sequences of admissible protocols
with a rate that is at least a fraction $r$ of the capacity at a
given SNR, i.e.,
\begin{equation}
E_{\NS}(r,\mrm{SNR}) = \sup_{\{\pi_K\}\in\cP(r,\mrm{SNR})}
E_{\NS}^-(\{\pi_K\}),\label{eq:NS-exp}
\end{equation}
where $\cP(r,\mrm{SNR})$ is a set of sequences of admissible
protocols with a rate that is a fraction $r$ of the capacity.
\end{defn}

The following establishes that, as with capacity, unicasting and
multicasting are not distinguished by their network scaling exponents.
\begin{prop}
The network scaling exponent is the same for both unicasting and
multicasting.
\end{prop}
\begin{proof}
First, for any sequence of admissible protocols,
\begin{equation*}
\Pr\{\cE^\mrm{mc}_{\pi_K}\} \geq \Pr\{\cE^\mrm{uc}_{\pi_K}\},
\end{equation*}
so that
\begin{equation}
E_{\NS}^\mrm{mc}(r,\mrm{SNR}) \leq E_{\NS}^\mrm{uc}(r,\mrm{SNR}).
\label{eq:a}
\end{equation}

Furthermore, if $\{\pi^*_K\}$ achieves the supremum for the
unicasting system, then from a simple application of the union bound
it follows that, for each $K$,
\begin{equation*}
\Pr\{\cE^\mrm{mc}_{\pi_K^*}\} \leq K
\Pr\{\cE^\mrm{uc}_{\pi_K^*}\},
\end{equation*}
and hence
\begin{equation}
E_{\NS}^\mrm{mc}(r,\mrm{SNR}) \geq
E_{\NS}^{-,\mrm{mc}}(\{\pi^*_K\}) =
E_{\NS}^\mrm{uc}(r,\mrm{SNR}).
\label{eq:b}
\end{equation}
Combining \eqref{eq:a} and \eqref{eq:b} we obtain
$E_{\NS}^\mrm{mc}(r,\mrm{SNR}) = E_{\NS}^\mrm{uc}(r,\mrm{SNR})$ as
desired.
\end{proof}

In the remainder of this section, we analyze a lower bound on the
network scaling exponent by optimizing over the class of the two-phase
protocols described in Section~\ref{sec:ucp}. For a fixed choice of
$\al$ and $\beta$, we can express the fraction of the capacity
achieved by the protocol as
\begin{equation}
r(\al,\beta,\mrm{SNR}) =
\frac{R(\alpha,\beta,\mathrm{SNR})}{C(\mathrm{SNR})},
\label{eq:f_defn}
\end{equation}
where we have made the dependency of both $R$ and $C$ in
\eqref{eq:Reff} and \eqref{eq:Capacity}, on the parameters of
interest explicit. We define the network reliability function of
the user cooperation protocol in Section~\ref{sec:ucp} as
\begin{equation}
E_{\NS}^-(r,\mathrm{SNR}) = \sup_{\al, \beta :
r(\al,\beta,\mrm{SNR}) \leq r}\left\{-\lim_{K\rightarrow \infty}
\frac{\ln \Pr\{\cE_{\al,\beta,K}\}}{K}\right\},
\label{eq:scaling-exponent}
\end{equation}
which constitutes a lower bound on $E_{\NS}(r,\mathrm{SNR})$ in
\eqref{eq:NS-exp}. Note that in the above definition, we have
constrained $\alpha$ and $\beta$ to be constants independent of
$K$.

\begin{figure}[tbp]
\begin{center}
\includegraphics[width=3.5in]{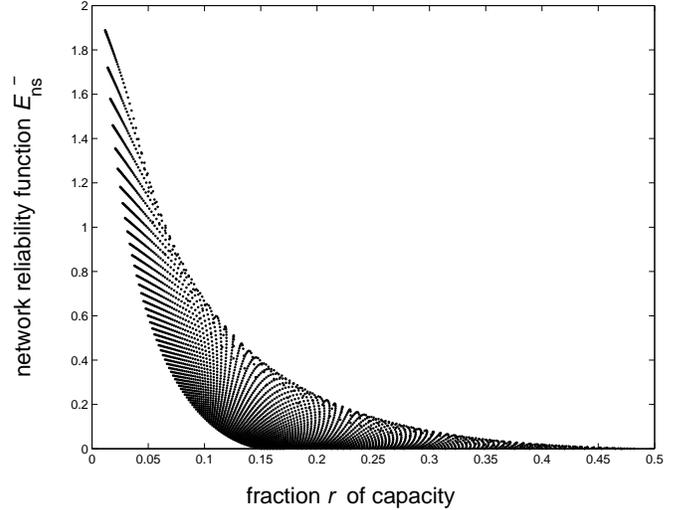}
\caption{The upper envelope of the plotted points indicates the
network scaling exponent for the two-phase cooperative protocol as a
function of the targetted fraction of capacity $r$.  Each point
corresponds to a particular value of $\alpha$ and $\beta$.  In this
example, $\mathrm{SNR} = 0$~dB.
\label{fig:Exponent}}
\end{center}
\end{figure}

The upper envelope of the points in Fig.~\ref{fig:Exponent}
indicates a lower bound on the network reliability function of our
two-phase protocol.  Each point in the plot corresponds to a
particular choice of $\alpha$ and $\beta$ in the protocol, for
which we have numerically evaluated $E_{\NS}^-$ in
\eqref{eq:scaling-exponent} for different values of $r$ at
$\mathrm{SNR} = 0$~dB.

Perhaps the most striking observation from Fig.~\ref{fig:Exponent} is
that the error exponent for the two-phase protocol is quite small when
aiming for rates that are more than about half of capacity.  This
implies that while the protocol is capacity achieving, it may require
a prohibitively large number of nodes to achieve rates anywhere close
to this capacity.  It remains to be determined whether there exist
more sophisticated protocols with substantially higher exponents in
this regime.

As a final comment, it should also be noted that
Fig.~\ref{fig:Exponent} effectively characterizes the efficient
operating frontier for the protocol. In particular, given a network
with $K$ nodes and an allowable outage probability $\epsilon$, one can
approximate the scaling exponent by $-\ln \epsilon/K$ and determine
the corresponding value of $r$, which is an estimate of how close one
can expect to get to capacity in the system.

\section{Concluding Remarks}
\label{sec:Conclusion and Future Work}

Perhaps the main contribution of this paper is a framework for
analyzing user cooperation protocols in the limit of large network
size (number of nodes), which we have illustrated in the case of a
multipath-rich Rayleigh fading environment.  Within this framework, we
have introduced a meaningful notion of Shannon capacity for this
regime and presented a simple two-phase protocol that can achieve
rates arbitrarily close to capacity.  A finer grain analysis of this
two-phase protocol in terms of its network scaling exponent, which
characterizes the rate of decay of error probabilty with network size,
shows that it may require prohibitively large number of nodes to
achieve rates close to the capacity with this protocol.

One important direction of future work is to study more sophisticated
models beyond the Rayleigh fading model within our framework.  One
could for example incorporate the effects of network geometry and
shadowing into the model. More generally, it would be of interest to
study a class of channel models for which user cooperation plays a
fundamental role in enabling reliable communication in
multicasting. The Rayleigh fading model considered here clearly
belongs to this class, but we believe the class may be quite rich and
may include many other models of practical importance.

Another important direction is to investigate how system performance
changes when the sum power constraint is replaced with individual
power constraints.  With individual power constraints, the system
capacity will increase with the number of nodes --- in fact, the MISO
upper bound increases according to $\Theta(\log K)$.  It remains to be
determined whether there exist cooperative multicasting protocols that
approach this upper bound or whether one can develop tighter upper
bounds for this scenario.

Finally, as noted in Section~\ref{Network-Scaling-Exponent}, the
two-phase protocol may require prohibitively large number of nodes
to achieve rates close to the capacity. It remains to investigate
whether more sophisticated protocols can improve the network
scaling exponent substantially in this regime.

\appendix[Derivation of Outage Approximation
  \protect\eqref{eq:unicast-Prob-approx}]

First, we write $\Pr\{\cE^\mathrm{uc}_{\al,\beta,K}\}$ in the form
\begin{align}
\Pr\{\cE^\mathrm{uc}_{\al,\beta,K}\} &= E\left[\Pr\{\cE^\mathrm{uc}_{\al,\beta,K}|K_1\}\right]\notag\\
 &= \sum_{k_1=1}^{K}
 \Pr\{K_1=k_1\}\,\Pr\{\cE^\mathrm{uc}_{\al,\beta,K}\mid K_1=k_1\} \notag\\
 &\approx \sum_{k_1=\al(1-\beta)K}^{K}
  \Pr\{K_1=k_1\}\,\Pr\{\cE^\mathrm{uc}_{\al,\beta,K}\mid K_1=k_1\}. \notag\\
\label{eq:foo}
\end{align}
Note that we have dropped the contribution of terms with $k_1\leq
\al(1-\beta)K$ in the summation, since we expect their aggregate
sum to be small as they deviate significantly from the mean
$E[K_1]= \al K$.

We now approximate each of the two factors in \eqref{eq:foo}. The
right factor we approximate by the upper bound
\eqref{eq:cond_Pr_single}.  The left factor we replace with via
Stirling's approximation for binomial distributions \cite[p.
284]{CoverBook}, yielding
\begin{align}
\Pr\{K_1=k_1\} &= \binom{K}{k_1} \al^{k_1} (1-\al)^{(K-k_1)}\notag\\
&\approx \frac{1}{\sqrt{K}}\exp\left\{-KD\biggl(\frac{k_1}{K}\biggm\|\al\biggr)\right\},
\label{eq:binom_Deviation}
\end{align}
where $D(\cdot\|\cdot)$ denotes the binary relative entropy function,
i.e., for any $0<p,q<1$,
\begin{equation}
D(p\|q) \defeq p\ln \frac{p}{q} + (1-p) \ln \frac{1-p}{1-q},
\end{equation}
and where $\alpha$ is the parameter of $K_1$
(cf.~\eqref{eq:K_1_defn}).

Thus, substituting \eqref{eq:cond_Pr_single} and
\eqref{eq:binom_Deviation} into \eqref{eq:foo} yields
\begin{align}
&\Pr\{\cE^\mathrm{uc}_{\al,\beta,K}\} \notag\\
&\approx \sum_{k_1=\al(1-\beta)K}^K \frac{1}{\sqrt{K}}\exp\left\{-KD\biggl(\frac{k_1}{K}\biggm\|\al\biggr)\right\} \quad \times  \notag\\
&\quad \quad \quad \left(\frac{\alpha K(1-\beta)e}{k_1}\right)^{k_1}
{\exp\left\{-\al K(1-\beta)\right\}}.
\label{eq:bar}
\end{align}
Finally, we approximate \eqref{eq:bar} by an approximation to the
largest single term in the summation, viz.,
\begin{multline}
\Pr\{\cE^\mathrm{uc}_{\al,\beta,K}\} \approx
\frac{1}{\sqrt{K}}{\exp\left\{-\al K(1-\beta)\right\}}
\quad \times \\
\qquad\max_{\gamma\in (\al(1-\beta),1)}
\exp\left\{-KD(\gamma\|\al)+ \gamma
K\ln\left(\frac{\al(1-\beta)e}{\gamma}\right)\right\}.
\label{eq:AppII-Bound0}
\end{multline}

Since the term in the exponent being minimized in
\eqref{eq:AppII-Bound0} is differentiable and convex in $\gamma$,
the optimizing $\gamma$ is the value at which the associated
derivative is zero, i.e.,
\begin{equation}
\frac{\gamma^2}{1-\gamma} = \mu
\label{eq:quadratic}
\end{equation}
where $\mu$ is as given in \eqref{eq:mu}.  Finally it is
straightforward to verify that \eqref{eq:quadratic} has a solution
in $(\al(1-\beta),1)$ and it may be solved explicitly, yielding
\eqref{eq:ga}.

\section*{Acknowledgements}

The authors thank the anonymous reviewers for several observations
which helped to improve the quality of the manuscript.

 \vfill

\end{document}